%% file: CSI-CLIP.tex
\documentclass[conference]{IEEEtran}
\pdfoutput=1
\IEEEoverridecommandlockouts
\usepackage{subfigure}
\usepackage{cite}
 \usepackage{siunitx}
\usepackage{amsmath,amssymb,amsfonts}
\usepackage{algorithmic}
\usepackage{algorithm}
\usepackage{multirow}
\usepackage{graphicx}
\usepackage{textcomp}
\usepackage{xcolor}
\usepackage{listings}
\usepackage{booktabs}
\usepackage{stfloats}
\usepackage{balance}

\def\BibTeX{{\rm B\kern-.05em{\sc i\kern-.025em b}\kern-.08em
    T\kern-.1667em\lower.7ex\hbox{E}\kern-.125emX}}
\begin{document}

\title{A MIMO Wireless Channel Foundation Model \\ via CIR-CSI Consistency

\author{Jun Jiang\IEEEauthorrefmark{1}, Wenjun Yu\IEEEauthorrefmark{1},Yunfan Li\IEEEauthorrefmark{1}, Yuan Gao\IEEEauthorrefmark{1}, and Shugong Xu\IEEEauthorrefmark{2}\IEEEauthorrefmark{3}\\
\IEEEauthorrefmark{1}School of Communication and Information Engineering, Shanghai University, Shanghai, China \\
\IEEEauthorrefmark{2}Xi'an Jiaotong-Liverpool University, Jiangsu, China \\
Email: \{jun\_jiang, yuwenjun, lyf2023, gaoyuansie\}@shu.edu.cn, shugong.xu@xjtlu.edu.cn \\
\IEEEauthorrefmark{3}Corresponding author
}
}
\maketitle

\begin{abstract}
In the field of artificial intelligence, self-supervised learning has demonstrated superior generalization capabilities by leveraging large-scale unlabeled datasets for pretraining, which is especially critical for wireless communication models to adapt to a variety of scenarios. This paper innovatively treats Channel State Information (CSI) and Channel Impulse Response (CIR) as naturally aligned multi-modal data and proposes the first MIMO wireless channel foundation model, named CSI-CLIP. By effectively capturing the joint representations of both CIR and CSI, CSI-CLIP exhibits remarkable adaptability across scenarios and robust feature extraction capabilities. Experimental results show that in positioning task, CSI-CLIP reduces the mean error distance by 22\%; in beam management task, it increases accuracy by 1\% compared to traditional supervised methods, as well as in the channel identification task. These improvements not only highlight the potential and value of CSI-CLIP in integrating sensing and communication but also demonstrate its significant advantages over existing techniques. Moreover, viewing CSI and CIR as multi-modal pairs and contrastive learning for wireless channel foundation model open up new research directions in the domain of MIMO wireless communications.
\end{abstract}


\section{Introduction}
Self-supervised learning (SSL) has emerged as a potent paradigm within machine learning, establishing a significant presence in the field of artificial intelligence\cite{bert,dino}. By uncovering intrinsic structures and feature representations from unannotated datasets, SSL enables models to acquire more generalized knowledge, which is especially critical for wireless communication systems. This capability showcases remarkable flexibility in diverse application scenarios, not only achieving superior performance on specific tasks but also enhancing system robustness and adaptability to tackle complex real-world environmental challenges. The foundation model pretrained with SSL demonstrates exceptional capability in extracting general feature representations, thereby providing substantial performance enhancements for various downstream tasks in wireless communication systems. It effectively transfers the general knowledge acquired to domain-specific applications, including but not limited to wireless localization, beam management (BM), and channel identification.

In wireless communication systems, channel identification stands as a pivotal step to ensure high-quality communication. It entails the capacity to differentiate between line-of-sight (LoS) and non-LoS (NLoS) conditions, directly impacting the effectiveness of dynamic spectrum management and power control, thus ensuring optimized resource allocation and reduced interference. Accurate channel recognition is fundamental for efficient and reliable communication services.

Positioning, as one of the core tasks in integrated sensing and communication (ISAC), holds obvious importance. Through precise positioning services, personalized user experiences such as accurate navigation or indoor location can be provided, while also offering critical data support for network optimization\cite{isacsurvey}. For instance, the high-precision positioning, imaging, and environmental reconstruction capabilities enabled by sensing contribute to improved communication performance, including but not limited to more accurate beamforming and faster beam failure recovery mechanisms\cite{isac}.

Furthermore, with the evolution of communication technologies, especially entering the era of 5th-Generation Mobile Communication Technology (5G) and beyond to 6th, beam management has become an essential component for efficient data transmission to achieve faster data rates and overcome the challenges of high-frequency channel fading\cite{bmsurvey}. BM aims to enhance signal strength within target areas by adjusting the direction of antenna arrays to form directional beams. However, the introduction of large-scale multi-input multi-output (MIMO) systems complicates the BM process and intensifies resource demands. In configurations featuring 32 transmitting antennas paired with 8 receiving antennas, a complete BM operation can take approximately 160 milliseconds, with required time increasing exponentially as the number of antennas grows\cite{sun2023ai}. To address this challenge, researchers are actively exploring possibilities to optimize the BM process using Artificial Intelligence (AI) technologies, aiming to boost efficiency and reduce costs.

It is noteworthy that the 3rd Generation Partnership Project (3GPP) has acknowledged the immense potential of AI in the realm of wireless communications, listing it among key application directions. Document TR 38.843\cite{3gpp.38.843} highlights positioning and beam management as primary scenes for AI empowerment, underscoring their strategic importance in the design of future communication architectures.

Against this backdrop, we innovatively regard channel impulse response (CIR) and channel state information (CSI) as multimodal data, a perspective shift that introduces new research directions within MIMO wireless communications. By incorporating contrastive learning into self-supervised pretraining methods\cite{clip}, the model can concurrently capture features from both modalities, thereby providing a more generalizable and efficient solution.

In summary, the main contributions of this paper can be summarized as follows.
\begin{enumerate}
    \item We introduce the first MIMO wireless channel foundation model, specifically designed for perception tasks, named CSI-CLIP. CSI-CLIP is capable of efficiently processing two distinct forms of data: CIR and CSI. Moreover, its design is task-agnostic, endowing it with versatile applicability across a multitude of scenarios.
    
    \item Furthermore, CSI-CLIP demonstrates outstanding cross-scenario adaptability and robust feature extraction capabilities. The model can capture and represent features within both CIR and CSI without relying on task-specific information, significantly enhancing its value as a general-purpose tool.
    
    \item The experimental results indicate that CSI-CLIP achieves substantial performance improvements compared to traditional supervised methods. In particular, in positioning tasks, there is an average performance increase of 22\%, while in beam management task, the accuracy has improved by 1\%, also improved in the channel identification task. These outcome not only highlight the potential and value of CSI-CLIP in practical applications but also underscore its superiority over existing approaches.
    
\end{enumerate}

\section{Related Works}
\label{sec:related works}

\subsection{Foundation Model in Wireless Communication}
The exploration of foundation models within the domain of wireless communication has recently gained significant momentum. This surge in attention is driven by the potential of foundation models to leverage self-supervised learning paradigms, such as Masked X Modeling (MXM), which facilitate seamless adaptation across various signal modalities.

Notably, Ott et al. \cite{ott2024radio} introduced an radio foundationm model for 5G indoor positioning through the innovative use of Masked Time-step Modeling. Meanwhile, LWM \cite{lwm} pioneered the development of the first channel-aware foundation model using Masked Channel Modeling, specifically channel identification and Sub-6G to mmWave beam prediction on the DeepMIMO dataset \cite{deepmimo}. Additionally, Aboulfotouh et al. \cite{MSM} advanced human activity sensing and spectrum segmentation within WiFi environments by employing Masked Spectrogram Modeling.

\begin{figure}[t]
\centering\includegraphics[width=1.05\columnwidth]{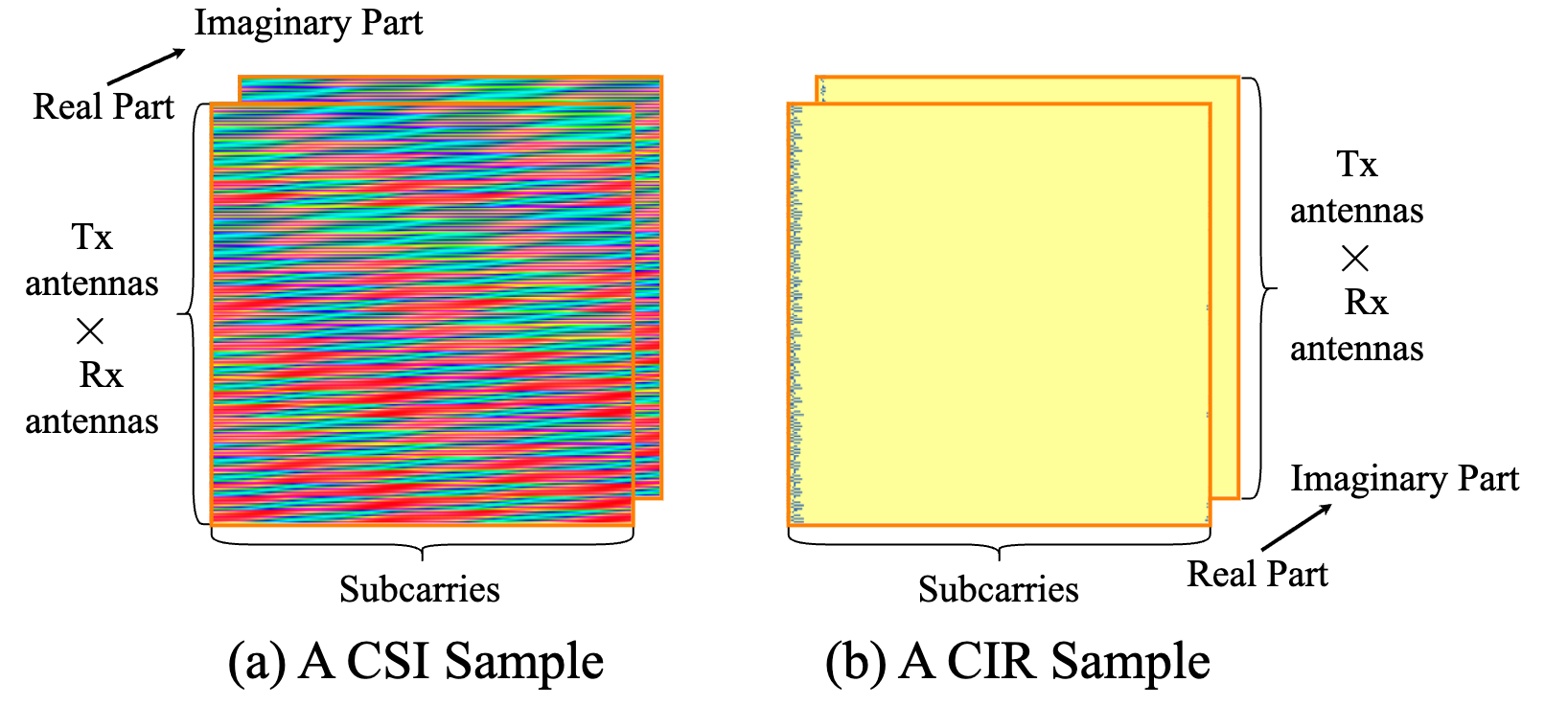}
\caption{Visualize for the CSI and CIR in MIMO topology.}
\label{fig:visual}
\end{figure}

Despite these advancements, existing literature predominantly centers on Multiple-Input Single-Output (MISO) systems, overlooking the complexities associated with MIMO setups. A critical challenge in MIMO systems involves handling CSI characterized by pronounced periodic patterns and sparse CIR data as shown in the Fig~\ref{fig:visual}. Current MXM pretraining methods may be inadequate in this context, as they tend to exploit strong correlations within unmasked data segments for straightforward signal reconstruction. Such operations, including interpolation or repetition, remain effective even under conditions of high masking ratios, thereby impeding the acquisition of more generalized feature representations necessary for robust MIMO processing.

\subsection{Contrastive Learning}
Contrastive learning, as a form of self-supervised learning, aims to learn data representations by minimizing the distance between positive sample pairs and maximizing the distance from negative sample pairs. This approach has achieved significant advancements in domains such as image and text processing, showcasing robust performance. Models like SimCLR\cite{simclr} and MoCo\cite{moco} underscore the importance of maintaining consistency in data representation across different views. They are trained by treating different views of the same data as positive pairs and other data instances as negative pairs.

However, the success of these methods largely hinges on effective data augmentation strategies. In the domain of images, techniques such as random cropping and color distortion have proven to be effective augmentation means. For fields like wireless communications, however, implementing contrastive learning becomes more complex due to the lack of standardized data augmentation practices. Moreover, multimodal models like CLIP\cite{clip}, while not relying on traditional data augmentation, require substantial paired multimodal data, which can be challenging to obtain in practical applications.

\begin{figure*}[tbp]
\centering\includegraphics[width=0.8\textwidth]{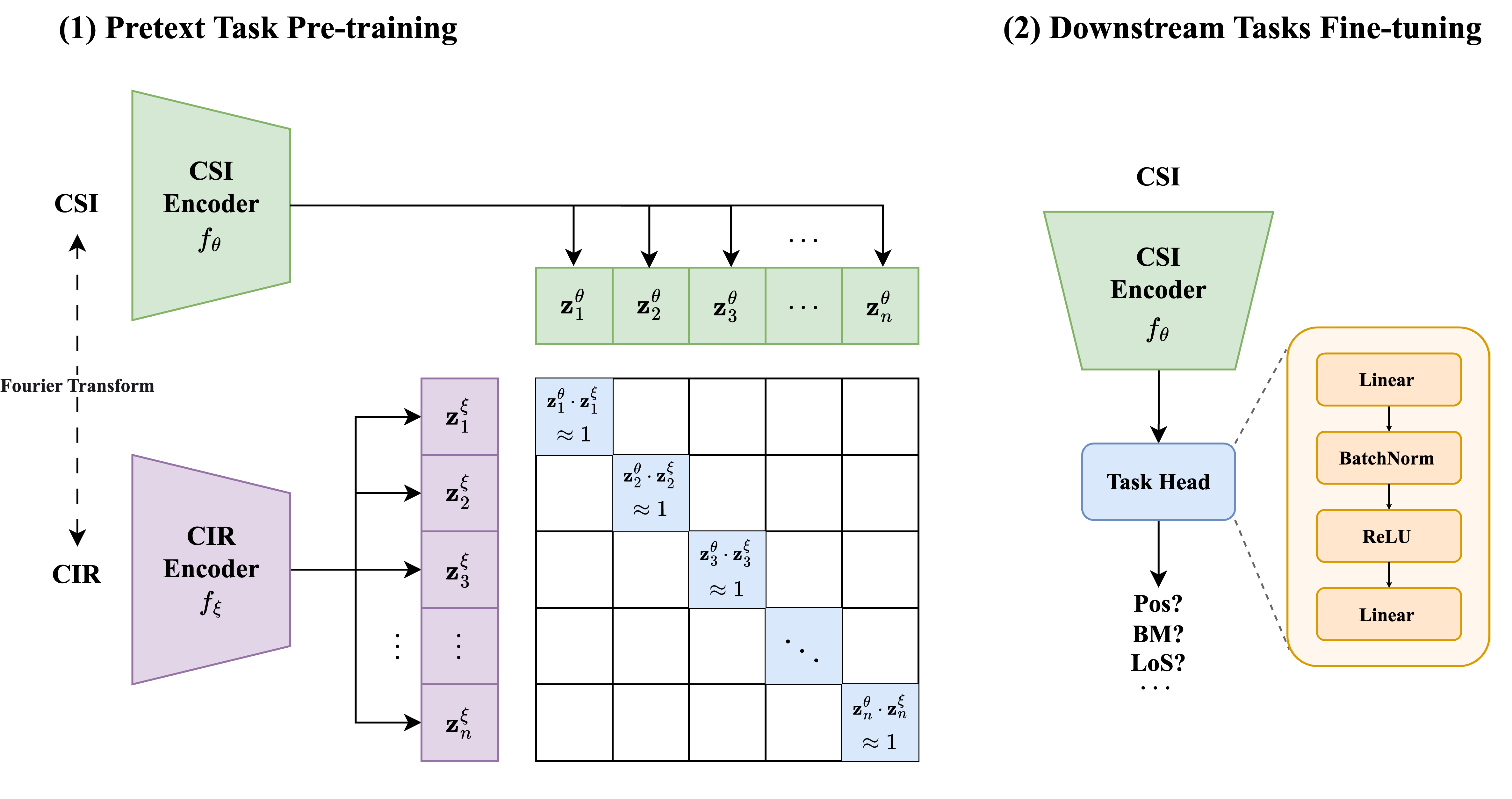}
\caption{Architecture of the proposed CSI-CLIP.}
\label{fig:pipeline}
\end{figure*}

In addition, TF-C\cite{tf-c}, a pioneering method in time series analysis, demonstrates that by combining temporal and frequency domain features through three contrastive learning approaches (time domain, frequency domain, and time-frequency domain), the performance of downstream tasks can be significantly enhanced. Specifically, it employs a consistency loss to minimize the distance between time-based and frequency-based embeddings, thereby enforcing consistency in the latent space between the two domains. This strategy not only improves the feature representation capabilities of pretrained models but also provides a robust framework for time series analysis.

\section{Proposed framework}
\label{sec:csi-clip}

Building on TF-C\cite{tf-c}, CSI-CLIP proposes a novel approach that views CIR and CSI as naturally aligned multimodal data pairs. Unlike TF-C, which uses three contrastive learning mechanisms, CSI-CLIP adopts a single contrastive learning framework, similar to the CLIP model in computer vision. The original time series and its Fourier-transformed spectrogram are considered as positive sample pairs, while all other pairs serve as negative samples for contrastive learning. This design is particularly well-suited for CIR and CSI in wireless communications, as both are inherently aligned in the time-frequency domain. Using multimodal contrastive learning, this approach captures features from both the time and frequency domains simultaneously, without the need for additional data augmentation, enhancing the feature representation capabilities of pretrained models.
\subsection{Pretext Task}
The architecture of CSI-CLIP draws inspiration from the recent advancements in contrastive learning within natural language processing and computer vision. As depicted in Fig.~\ref{fig:pipeline}, the proposed model features a dual-pathway design, where each pathway is meticulously tailored for the processing of CIR and CSI, two modalities that, while distinct, share a close relationship within the domain of wireless communication data. The congruity of the backbone network structure across both pathways ensures that the extracted features possess a broad generalization capability while still capturing the idiosyncrasies unique to each modality.

During the feature extraction stage, each pathway operates autonomously, adapting its operations to the intrinsic properties of its input. For instance, CIR data are characterized by temporal information on signal propagation paths, whereas CSI encapsulates characteristics within the frequency domain. CSI-CLIP enhances its ability to uncover latent patterns within the data. The outputs from the two pathways are subsequently mapped into a shared embedding space, allowing for a unified representation of CIR and CSI.

\begin{equation}
\begin{split}
    \mathbf{z}^{\theta}_{i} = f_\theta(CSI_i)\\
    \mathbf{z}_{\xi} = f_\xi(CIR_i)
\end{split}
\end{equation}

where $f_\theta$ and $f_\xi$ denote the encoder for CSI and CIR, respectively, and $\mathbf{z}^{\theta}_i$ and $\mathbf{z}^{\xi}_i$ represent the embeddings of CSI and CIR for the  $i^{\text{th}}$ sample.

To ensure that the embeddings of the two modalities maintain a meaningful relationship, a contrastive learning objective is introduced. This objective function aims to minimize the distance between positive pairs (embeddings of the same sample from different modalities) and maximize the distance between negative pairs (embeddings of different samples). Specifically, the cosine similarity metric is employed to measure the alignment of the embeddings. The learnable temperature parameter $\tau$ controls the sharpness of the similarity distribution, allowing more nuanced control over contrast loss.

\begin{equation}
\mathcal{L} = -\frac{1}{N} \sum_{i=1}^{N} \log \frac{\exp(\cos(\mathbf{z}_i^\theta, \mathbf{z}_i^\xi) / \tau)}{\sum_{j=1}^{N} \exp(\cos(\mathbf{z}_i^\theta, \mathbf{z}_j^\xi) / \tau)}
\label{equ:loss}
\end{equation}

Here, $N$ represents the batch size, and $\cos(\cdot, \cdot)$ denotes the cosine similarity between two vectors. By optimizing this loss function, the model learns to produce embeddings that preserve the intrinsic relationships between CIR and CSI representations.

\subsection{Downstream Tasks}
In the downstream task fine-tuning stage, transfer learning is used to leverage the rich feature representations embedded within pretrained models to boost task performance. Specifically, the pretrained CSI Encoder $f_\theta$ has already acquired the ability to extract general features from CSI data. These features exhibit broad applicability across a variety of CSI-based applications. By appending a task-specific head composed of two fully connected layers and subsequently fine-tuning this architecture with labeled data, the model can optimize the model for particular downstream tasks.

\subsubsection{Channel Identification}
Channel Identification task can be formulated as a typical binary classification task, where the objective is to classify the input CSI into categories such as LoS and NLoS. The model is trained using the Cross-Entropy loss function as follow.
\begin{equation}\label{equ:cross_entropy}
    \mathcal{L} = -\sum_{c=1}^{C} y_c \log(\hat{y}_c),
\end{equation}
where $C$ is the number of classes, $y_c$ is the indicator variable that equals one if class $c$ is the true class for the sample and zero otherwise, and $\hat{y}_c$ is the predicted probability that the sample belongs to class $c$.
\subsubsection{Positioning}
Positioning can be formulated as a regression task aimed at estimating the coordinates of a user or device within a given space. The goal is to predict continuous values representing the location based on CSI. 

The training process involves minimizing a Mean Squared Error (MSE) loss function, which quantifies the average squared difference between the predicted and actual positions.
\begin{equation}\label{equ:mse}
\mathcal{L} = \frac{1}{N} \sum_{i=1}^{N}  \left\| \mathbf{p}_i -\hat{\mathbf{p}}_i \right\|^2 
\end{equation}

where $\mathbf{p}_i$ represents the true position coordinates of the $i^{\text{th}}$ sample, $\hat{\mathbf{p}}_i$ denotes the predicted position, and $N$ is the total number of samples.

\subsubsection{Beam Management}
The goal of BM is to predict the optimal beam index $b^*$ by using the CSI of the user, denoted as $\mathbf{H}$, and projecting it onto the steering vector $s(b)$ for each beam in the codebook. The received power for a given beam $b$ can be calculated as $P_{\mathrm{rx}}(b) = \| \mathbf{H} \cdot s(b) \|^2$.

Consequently, the optimal beam index $b^*$ is selected based on the maximization of the received power.
\begin{equation}
    b^* = \underset{b}{\operatorname{argmax}} \, P_{\mathrm{rx}}(b)
\end{equation}

This optimization problem can be reformulated as a multiclass classification task, where the objective is to classify the optimal beam index $b^*$. The model is trained using the Cross-Entropy loss function, as defined in equation~\ref{equ:cross_entropy}.

\section{Experiment}
\label{sec:experiment}

\subsection{Dataset}
The CSI-CLIP model utilizes an extensive and varied dataset, which is integral to its design. This dataset encompasses over 700,000 CSI samples, sourced from 35 distinct scenarios within the DeepMIMO dataset\cite{deepmimo}. The diversity of these scenarios, both in terms of environment type and operating frequency, facilitates robust generalization across a multitude of wireless communication settings.

The dataset includes 35 different scenarios, covering a broad spectrum of indoor and outdoor environments. Operating frequencies range from Sub-6 GHz through mmWave up to terahertz, thus offering a comprehensive overview of the spectral bands relevant to contemporary wireless communications.

To maintain balanced representation across scenarios, a stratified sampling method is employed. Scenarios with fewer than 50,000 users are fully included, whereas for those exceeding this threshold, a subset is selected to preserve diversity and prevent any single scenario from disproportionately influencing the dataset. This approach guarantees exposure to a wide array of conditions and user distributions, thereby enhancing the model's robustness and generalization.

Before being input into the model, all CSI samples are preprocessed, including min-max normalization and standardization, to ensure consistency and improve learning efficiency. During the fine-tuning stage for downstream tasks, 80\% of the data is allocated for training, and others are reserved for validation. For BM task, 64 beams DFT codebook is used.

To demonstrate the generalization capability of CSI-CLIP, additional data were generated using Sionna RT \cite{Sionna} under identical configurations. Specifically, the etoile urban cellular communication scenario provided by Sionna was selected for this purpose. In this setup, User Equipment (UE) nodes are distributed at intervals of 1 meter within a typical single-site, three-sector layout, with the maximum distance to the Base Station (BS) not exceeding 200 meters. The BS is positioned at coordinates (0,0,30), while the UEs are located within the three sectors at a height of 1.5 meters. The carrier frequency is set to 3.5 GHz, and the maximum number of interactions between rays and scene objects is limited to 4.

\input{table}

\subsection{Implementation Details}
In our implementation, the simulation parameters are configured as follows: the maximum number of propagation paths is 20; the system bandwidth is 10 MHz, and the number of subcarriers is 256. The base station employs an 8 $\times$ 8 uniform planar array (UPA), whereas user terminals feature a 2 $\times$ 2 UPA configuration.

The encoder architecture is based on the ResNet50\cite{resnet} model, adapted to accommodate a two-channel input. All phases of training and testing for CSI-CLIP were executed on NVIDIA GeForce RTX 4090 GPUs. A batch size of 128 was adopted, and the training process was carried out for up to 300 epochs, with early stopping criteria applied to mitigate overfitting. An AdamW optimizer was used, with an initial learning rate of 0.0008, which is reduced by 20\% every ten epochs if no improvement in minimal validation loss.

\subsection{Results analysis}
The comparative analysis in Table~\ref{tab:result} demonstrates CSI-CLIP's superior performance over supervised baselines in positioning and BM tasks across 35 heterogeneous scenarios. Despite using identical encoder structures and task heads, the key difference lies in their training paradigms. Supervised models are trained on individual scenario datasets without cross-modal alignment between CSI and CIR representations. In contrast, the pretrained encoder in CSI-CLIP aligns features from both CSI and CIR, enabling it to capture their shared characteristics. This alignment significantly enhances generalization across diverse environments and frequency bands.

In urban settings, CSI-CLIP achieves substantial reductions in positioning error and improvements in BM accuracy. For example, in Los Angeles, it reduced the average error distance from 49.03 meters to 34.18 meters (a decrease of 30.29\%) and increased the precision of the BM from 75.68\% to 78.38\% (a gain of 2.70\%). Notably, in scenarios with limited data, CSI-CLIP outperforms baselines, highlighting its practical value.

Even in challenging environments, such as the O1 drone scenario at 200GHz, CSI-CLIP improved positioning by 40.02\% and BM accuracy by 0.35\%. These outcomes confirm that CSI-CLIP not only maintains high efficiency within specific cities or frequency ranges but also consistently performs well under a broad set of conditions, demonstrating stability.

In the positioning task, despite the significant differences in error distances, CSI-CLIP generally achieved improvements ranging from 10\% to 40\%, average 22\%. For the BM task, while the degree of improvement was relatively smaller, average 1\%, CSI-CLIP provided positive improvements in almost all scenarios tested, reflecting consistent effectiveness. However, in some scenarios such as I2, a drop was observed, possibly due to the unique scenario characteristics.

\begin{table}[tbp]
\centering
\caption{Performance Comparison on Channel Identification Task.}
\resizebox{\columnwidth}{!}{%
\begin{tabular}{@{}cccc@{}}
\toprule
Scenario      & Supervised Learning & CSI-CLIP(Ours) & Improvemnet     \\ \midrule
Boston5G\_28G & 99.68\%             & 100.00\%       & \textbf{0.32\%} \\
city\_phoenix & 100.00\%            & 100.00\%       & \textbf{0.00\%} \\
I1\_2.5G      & 100.00\%            & 100.00\%       & \textbf{0.00\%} \\
O1\_140G      & 99.97\%             & 99.99\%        & \textbf{0.02\%} \\
officefloor   & 97.36\%             & 98.62\%        & \textbf{1.26\%} \\ \bottomrule
\end{tabular}%
}

\label{tab:los}
\end{table}

Table~\ref{tab:los} showcases the performance comparison of the channel identification task across five distinct scenarios. Although channel identification is a binary classification problem in which features under LoS and NLoS are relatively distinguishable, leading to high accuracy (almost 100\%) in most scenarios, CSI-CLIP still demonstrates exceptional capability.

CSI-CLIP not only achieves extremely high accuracy in the majority of test scenarios but also realizes additional performance improvements over already excellent baselines. This achievement underscores the effectiveness and potential of CSI-CLIP in channel identification task, particularly in scenarios where features are less pronounced or environmental conditions are more complex. 

\begin{table}[tbp]
\centering
\caption{Performance on Dataset Generated by Sionna for Positioning.}
\resizebox{\columnwidth}{!}{%
\begin{tabular}{@{}cccc@{}}
\toprule
Sector & Supervised Learning & CSI-CLIP(Ours) & Improvemnet      \\ \midrule
A        & 2.71                & 0.93           & \textbf{65.68\%} \\
B        & 5.57                & 4.25           & \textbf{23.70\%} \\
C        & 37.50               & 30.45          & \textbf{18.80\%} \\ \bottomrule
\end{tabular}%
}
\label{tab:sionna}
\end{table}

Furthermore, Table~\ref{tab:sionna} illustrates the performance of CSI-CLIP on positioning task based on the SionnaRT simulation dataset, where the model was initially pretrained on the DeepMIMO dataset. The results indicate that CSI-CLIP exhibits robust generalization capabilities even when confronted with data not encountered during training, achieving significant performance improvements compared to traditional supervised learning methods. Not only does CSI-CLIP excel on known datasets, but it also maintains a high level of accuracy and reliability in new environments. By offering more refined CSI embeddings, CSI-CLIP can make important contributions to the further optimization of ISAC systems.

\section{Conclusions and Limitations}
\label{sec:conclusion}

This paper introduces the first MIMO wireless channels foundational model, named  CSI-CLIP. It offers a comprehensive solution by simultaneously capturing the joint feature representation of CSI and CIR. Consequently, CSI-CLIP significantly enhances performance across multiple downstream tasks. Experimental results demonstrate that compared to traditional supervised training methods, CSI-CLIP exhibits more robust feature representation and stronger generalization when processing CSI under different scenarios and frequencies, presenting substantial application prospects for practical ISAC systems.

Nevertheless, CSI-CLIP has certain limitations. For instance, it does not generalize well to scenarios with differing numbers of transmit/receive antennas and subcarriers from those used in our training configurations. The extracted CSI feature representations are not strong enough to support linear-probe. It is worth noting that previous studies have also struggled with these two issues, making them important directions for future research.

\bibliographystyle{IEEEtran}
\bibliography{bibfile.bbl}
\balance

\end{document}

%% file: table.tex
\begin{table*}[]
\centering
\caption{Performance Comparison between CSI-CLIP and Supervised Models.We use mean error distance(m) to evulate the positioning task and accuary for BM task.}
\resizebox{0.75\textwidth}{!}{%
\begin{tabular}{@{}ccccccccc@{}}
\toprule
 &
  Scenario &
  \multicolumn{7}{c}{city} \\
 &
  City Name &
  newyork &
  losangeles &
  chicago &
  houston &
  phoenix &
  philadelphia &
  miami \\
 &
  Train Samples &
  1026 &
  592 &
  228 &
  2060 &
  2163 &
  503 &
  1345 \\
\multirow{-4}{*}{} &
  Validation Samples &
  257 &
  148 &
  57 &
  515 &
  541 &
  126 &
  337 \\ \midrule
\multicolumn{1}{c|}{} &
  Supervised Learning &
  41.19 &
  49.03 &
  305.4 &
  66.53 &
  10.65 &
  34.1 &
  56.29 \\
\multicolumn{1}{c|}{} &
  CSI-CLIP(Ours) &
  36.15 &
  34.18 &
  241.4 &
  46.67 &
  9.34 &
  19.91 &
  44.18 \\
\multicolumn{1}{c|}{\multirow{-3}{*}{Positioning}} &
  \textbf{Improvement} &
  {\color[HTML]{FF0000} \textbf{12.24\%}} &
  {\color[HTML]{FF0000} \textbf{30.29\%}} &
  {\color[HTML]{FF0000} \textbf{20.96\%}} &
  {\color[HTML]{FF0000} \textbf{29.85\%}} &
  {\color[HTML]{FF0000} \textbf{12.30\%}} &
  {\color[HTML]{FF0000} \textbf{41.61\%}} &
  {\color[HTML]{FF0000} \textbf{21.51\%}} \\ \midrule
\multicolumn{1}{c|}{} &
  Supervised Learning &
  75.10\% &
  75.68\% &
  87.72\% &
  80.78\% &
  82.07\% &
  79.37\% &
  81.60\% \\
\multicolumn{1}{c|}{} &
  CSI-CLIP(Ours) &
  77.04\% &
  78.38\% &
  89.47\% &
  80.97\% &
  82.99\% &
  80.95\% &
  82.20\% \\
\multicolumn{1}{c|}{\multirow{-3}{*}{Beam Management}} &
  \textbf{Improvement} &
  {\color[HTML]{0070C0} \textbf{1.94\%}} &
  {\color[HTML]{0070C0} \textbf{2.70\%}} &
  {\color[HTML]{0070C0} \textbf{1.75\%}} &
  {\color[HTML]{0070C0} \textbf{0.19\%}} &
  {\color[HTML]{0070C0} \textbf{0.92\%}} &
  {\color[HTML]{0070C0} \textbf{1.58\%}} &
  {\color[HTML]{0070C0} \textbf{0.60\%}} \\ \midrule
 &
  Scenario &
  \multicolumn{7}{c}{city} \\
 &
  City Name &
  sandiego &
  dallas &
  sanfrancisco &
  austin &
  santaclara &
  fortworth &
  columbus \\
 &
  Train Samples &
  1753 &
  2005 &
  1326 &
  1482 &
  2151 &
  1521 &
  1148 \\
\multirow{-4}{*}{} &
  Validation Samples &
  439 &
  502 &
  332 &
  371 &
  538 &
  381 &
  288 \\ \midrule
\multicolumn{1}{c|}{} &
  Supervised Learning &
  66.77 &
  737.2 &
  19.78 &
  14.16 &
  188.7 &
  171.4 &
  17.6 \\
\multicolumn{1}{c|}{} &
  CSI-CLIP(Ours) &
  56.65 &
  658.3 &
  14.17 &
  8.95 &
  165.2 &
  171.1 &
  14.48 \\
\multicolumn{1}{c|}{\multirow{-3}{*}{Positioning}} &
  \textbf{Improvement} &
  {\color[HTML]{FF0000} \textbf{15.16\%}} &
  {\color[HTML]{FF0000} \textbf{10.70\%}} &
  {\color[HTML]{FF0000} \textbf{28.36\%}} &
  {\color[HTML]{FF0000} \textbf{36.79\%}} &
  {\color[HTML]{FF0000} \textbf{12.45\%}} &
  {\color[HTML]{FF0000} \textbf{0.18\%}} &
  {\color[HTML]{FF0000} \textbf{17.73\%}} \\ \midrule
\multicolumn{1}{c|}{} &
  Supervised Learning &
  73.80\% &
  81.87\% &
  88.86\% &
  78.98\% &
  76.95\% &
  82.94\% &
  62.85\% \\
\multicolumn{1}{c|}{} &
  CSI-CLIP(Ours) &
  74.49\% &
  83.07\% &
  90.66\% &
  80.05\% &
  77.32\% &
  84.78\% &
  65.63\% \\
\multicolumn{1}{c|}{\multirow{-3}{*}{Beam Management}} &
  \textbf{Improvement} &
  {\color[HTML]{0070C0} \textbf{0.69\%}} &
  {\color[HTML]{0070C0} \textbf{1.20\%}} &
  {\color[HTML]{0070C0} \textbf{1.80\%}} &
  {\color[HTML]{0070C0} \textbf{1.07\%}} &
  {\color[HTML]{0070C0} \textbf{0.37\%}} &
  {\color[HTML]{0070C0} \textbf{1.84\%}} &
  {\color[HTML]{0070C0} \textbf{2.78\%}} \\ \midrule
 &
  Scenario &
  \multicolumn{6}{c}{city} &
  officefloor1 \\
 &
  City Name &
  charlotte &
  indianapolis &
  sanfrancisco &
  seattle &
  denver &
  oklahoma &
  / \\
 &
  Train Samples &
  2881 &
  2720 &
  2626 &
  1177 &
  832 &
  2764 &
  17541 \\
\multirow{-4}{*}{} &
  Validation Samples &
  721 &
  439 &
  657 &
  295 &
  209 &
  691 &
  3936 \\ \midrule
\multicolumn{1}{c|}{} &
  Supervised Learning &
  158.1 &
  48.08 &
  10.81 &
  39.88 &
  324.9 &
  18.87 &
  3.21 \\
\multicolumn{1}{c|}{} &
  CSI-CLIP(Ours) &
  139.9 &
  46.39 &
  5.71 &
  19.28 &
  266.7 &
  16.74 &
  3.16 \\
\multicolumn{1}{c|}{\multirow{-3}{*}{Positioning}} &
  \textbf{Improvement} &
  {\color[HTML]{FF0000} \textbf{11.51\%}} &
  {\color[HTML]{FF0000} \textbf{3.51\%}} &
  {\color[HTML]{FF0000} \textbf{47.18\%}} &
  {\color[HTML]{FF0000} \textbf{51.65\%}} &
  {\color[HTML]{FF0000} \textbf{17.91\%}} &
  {\color[HTML]{FF0000} \textbf{11.29\%}} &
  {\color[HTML]{FF0000} \textbf{1.56\%}} \\ \midrule
\multicolumn{1}{c|}{} &
  Supervised Learning &
  82.52\% &
  77.53\% &
  82.19\% &
  70.17\% &
  73.21\% &
  83.07\% &
  71.85\% \\
\multicolumn{1}{c|}{} &
  CSI-CLIP(Ours) &
  84.60\% &
  80.18\% &
  82.34\% &
  70.85\% &
  75.60\% &
  83.36\% &
  72.51\% \\
\multicolumn{1}{c|}{\multirow{-3}{*}{Beam Management}} &
  \textbf{Improvement} &
  {\color[HTML]{0070C0} \textbf{2.08\%}} &
  {\color[HTML]{0070C0} \textbf{2.65\%}} &
  {\color[HTML]{0070C0} \textbf{0.15\%}} &
  {\color[HTML]{0070C0} \textbf{0.68\%}} &
  {\color[HTML]{0070C0} \textbf{2.39\%}} &
  {\color[HTML]{0070C0} \textbf{0.29\%}} &
  {\color[HTML]{0070C0} \textbf{0.66\%}} \\ \midrule
 &
  Scenario &
  \multicolumn{2}{c}{Boston5G} &
  \multicolumn{2}{c}{I1} &
  I2\_B &
  I3 &
  O1\_drone \\
 &
  Frequency &
  3.5G &
  28G &
  2.4G &
  2.5G &
  28G &
  2.4G &
  200G \\
 &
  Train Samples &
  8741 &
  8732 &
  15693 &
  15693 &
  15702 &
  16552 &
  24362 \\
\multirow{-4}{*}{} &
  Validation Samples &
  2186 &
  2184 &
  3924 &
  3924 &
  3926 &
  4139 &
  6091 \\ \midrule
\multicolumn{1}{c|}{} &
  Supervised Learning &
  49.03 &
  30.38 &
  5.71E-03 &
  2.34E-03 &
  6.34E-02 &
  4.58E-03 &
  2.31 \\
\multicolumn{1}{c|}{} &
  CSI-CLIP(Ours) &
  35.68 &
  25.54 &
  3.31E-03 &
  2.35E-03 &
  6.54E-02 &
  2.89E-03 &
  1.39 \\
\multicolumn{1}{c|}{\multirow{-3}{*}{Positioning}} &
  \textbf{Improvement} &
  {\color[HTML]{FF0000} \textbf{27.23\%}} &
  {\color[HTML]{FF0000} \textbf{15.93\%}} &
  {\color[HTML]{FF0000} \textbf{42.03\%}} &
  {\color[HTML]{FF0000} \textbf{-0.43\%}} &
  {\color[HTML]{FF0000} \textbf{-3.15\%}} &
  {\color[HTML]{FF0000} \textbf{36.90\%}} &
  {\color[HTML]{FF0000} \textbf{39.83\%}} \\ \midrule
\multicolumn{1}{c|}{} &
  Supervised Learning &
  83.21\% &
  81.13\% &
  93.96\% &
  93.58\% &
  84.31\% &
  84.95\% &
  92.87\% \\
\multicolumn{1}{c|}{} &
  CSI-CLIP(Ours) &
  84.35\% &
  81.91\% &
  94.27\% &
  93.91\% &
  84.44\% &
  85.17\% &
  93.22\% \\
\multicolumn{1}{c|}{\multirow{-3}{*}{Beam Management}} &
  \textbf{Improvement} &
  {\color[HTML]{0070C0} \textbf{1.14\%}} &
  {\color[HTML]{0070C0} \textbf{0.78\%}} &
  {\color[HTML]{0070C0} \textbf{0.31\%}} &
  {\color[HTML]{0070C0} \textbf{0.33\%}} &
  {\color[HTML]{0070C0} \textbf{0.13\%}} &
  {\color[HTML]{0070C0} \textbf{0.22\%}} &
  {\color[HTML]{0070C0} \textbf{0.35\%}} \\ \midrule
 &
  Scenario &
  \multicolumn{5}{c}{O1} &
  \multicolumn{2}{c}{O1\_B} \\
 &
  Frequency &
  3.4G &
  3.5G &
  28G &
  60G &
  140G &
  3.5G &
  28G \\
 &
  Train Samples &
  \multicolumn{5}{c}{67017} &
  38712 &
  38585 \\
\multirow{-4}{*}{} &
  Validation Samples &
  \multicolumn{5}{c}{16755} &
  9679 &
  9647 \\ \midrule
\multicolumn{1}{c|}{} &
  Supervised Learning &
  1.26 &
  1.45 &
  5.27 &
  2.4 &
  3.18 &
  8.26 &
  24.15 \\
\multicolumn{1}{c|}{} &
  CSI-CLIP(Ours) &
  0.98 &
  1.22 &
  2.81 &
  1.74 &
  2.28 &
  6.3 &
  21.39 \\
\multicolumn{1}{c|}{\multirow{-3}{*}{Positioning}} &
  \textbf{Improvement} &
  {\color[HTML]{FF0000} \textbf{22.22\%}} &
  {\color[HTML]{FF0000} \textbf{15.86\%}} &
  {\color[HTML]{FF0000} \textbf{46.68\%}} &
  {\color[HTML]{FF0000} \textbf{27.50\%}} &
  {\color[HTML]{FF0000} \textbf{28.30\%}} &
  {\color[HTML]{FF0000} \textbf{23.73\%}} &
  {\color[HTML]{FF0000} \textbf{11.43\%}} \\ \midrule
\multicolumn{1}{c|}{} &
  Supervised Learning &
  95.92\% &
  95.75\% &
  95.27\% &
  95.04\% &
  94.87\% &
  93.70\% &
  92.27\% \\
\multicolumn{1}{c|}{} &
  CSI-CLIP(Ours) &
  95.94\% &
  95.95\% &
  95.33\% &
  95.23\% &
  94.91\% &
  93.89\% &
  92.68\% \\
\multicolumn{1}{c|}{\multirow{-3}{*}{Beam Management}} &
  \textbf{Improvement} &
  {\color[HTML]{0070C0} \textbf{0.02\%}} &
  {\color[HTML]{0070C0} \textbf{0.20\%}} &
  {\color[HTML]{0070C0} \textbf{0.06\%}} &
  {\color[HTML]{0070C0} \textbf{0.19\%}} &
  {\color[HTML]{0070C0} \textbf{0.04\%}} &
  {\color[HTML]{0070C0} \textbf{0.19\%}} &
  {\color[HTML]{0070C0} \textbf{0.41\%}} \\ \bottomrule
\end{tabular}%
}

\label{tab:result}
\end{table*}